\documentclass[twocolumn,amsmath,nofootinbib]{revtex4} 
\input{epsf.sty}

\def\rn{}
\def\nn#1 #2{#2. #1}				
\def\nnn#1 #2 #3{#2. #3. #1}			
\def\nnnn#1 #2 #3 #4{#2. #3. #4 #1}		
\def\nnnnn#1 #2 #3 #4 #5{#2. #3. #4 #5. #1}	
\def\dualand{ and\hbox{ }}				
\def\multiand{, and\hbox{ }}				
\def\rf#1;#2;#3;#4;#5 {{\frenchspacing\par\rn#1, #3 {\bf #4}, #5 (#2). \par}}
\def\rg#1;#2;#3;#4;#5;#6 {{\frenchspacing\par\rn#1, #3 {\bf #4}, #5 (#2). \par}}
\def\rfbook#1;#2;#3;#4;#5 {{\frenchspacing\par\rn#1, {\it #3} (#5, #4, #2).\par}}
\def\rfprep#1;#2;#3 {{\par\frenchspacing\rn#1, #3 (#2).\par}}
\def\rfproc#1;#2;#3;#4;#5;#6 {{\frenchspacing\par\rn#1 #2, in {\it #3}, ed. #4 (#5: #6)\par}}
\def\rfprocp#1;#2;#3;#4;#5;#6;#7 {{\frenchspacing\par\rn#1 #2, in {\it #3}, ed. #4 (#5: #6), p#7\par}}

\def\beq#1{\begin{equation}\label{#1}}
\def\eeq{\end{equation}}
\def\beqa#1{\begin{eqnarray}\label{#1}}
\def\eeqa{\end{eqnarray}}
\def\eq#1{equation~(\ref{#1})}
\def\Eq#1{Equation~(\ref{#1})}

\def\fig#1{Figure~\ref{#1}}
\def\Fig#1{Figure~\ref{#1}}

\def\spose#1{\hbox to 0pt{#1\hss}}
\def\simlt{\mathrel{\spose{\lower 3pt\hbox{$\mathchar"218$}}
     \raise 2.0pt\hbox{$\mathchar"13C$}}}
\def\simgt{\mathrel{\spose{\lower 3pt\hbox{$\mathchar"218$}}
     \raise 2.0pt\hbox{$\mathchar"13E$}}}
\def\simpropto{\mathrel{\spose{\lower 3pt\hbox{$\mathchar"218$}}
     \raise 2.0pt\hbox{$\propto$}}}

\def\etal{{\frenchspacing\it et al.}}
\def\ie{{\frenchspacing\it i.e.}}
\def\eg{{\frenchspacing\it e.g.}}
\def\etc{{\frenchspacing\it etc.}}

\def\fn{f_\nu}
\def\mn{m_\nu}
\def\mbc{m_{\rm bc}}
\def\rhol{\rho_\Lambda}
\def\rhom{\rho_{\rm m}}
\def\rhon{\rho_\nu}
\def\rhobc{\rho_{\rm bc}}
\def\Ol{\Omega_\Lambda}
\def\Om{\Omega_{\rm m}}

\def\Avac{A_\Lambda}
\def\xeq{x_{\rm eq}}
\def\zeq{z_{\rm eq}}
\def\Gl{G_\Lambda}
\def\Gg{G_\gamma}
\def\G{G}
\def\Ms{M_\odot}

\def\erfc{{\rm erfc}\,}

\newcommand{\beqq}{\begin{equation}}

\def\lap{\lower.5ex\hbox{$\; \buildrel < \over \sim \;$}}
\def\gap{\lower.5ex\hbox{$\; \buildrel > \over \sim \;$}}

\def\L{\Lambda}

\def\sigmaM{\sigma_M}

\begin{document}

\title{Anthropic predictions for neutrino masses} 

\author{Max Tegmark$^{1,2}$, Alexander Vilenkin$^3$ and Levon Pogosian$^3$}

\address{
$^1$ Dept. of Physics, Massachusetts Institute of Technology, Cambridge, MA 02139, USA\\
$^2$ Department of Physics and Astronomy, University of Pennsylvania, Philadelphia, PA 19104, USA\\
$^3$ Institute of Cosmology, Department of Physics and Astronomy, Tufts University, Medford, MA 02155, USA
}

\date{May 7 2005, Phys.~Rev.~D, in press.}

\begin{abstract}

It is argued that small values of the neutrino masses may be due to
anthropic selection effects. If this is the case, then
the combined mass of the three neutrino
species is expected to be $\sim 1$~eV, neutrinos causing
a non-negligible suppression of galaxy formation.

\end{abstract}
\bigskip

\maketitle

\section{Introduction}
 
The major ingredients of the Universe are dark energy, $\Omega_\Lambda
\sim 0.7$, and non-relativistic matter, $\Omega_m\sim 0.3$. The latter
consists of non-baryonic dark matter, $\Omega_D\sim 0.25$, baryons,
$\Omega_B\sim 0.05$, and massive neutrinos, $\Omega_\nu\gtrsim 0.001$.
The fact that $\Omega_\Lambda$ is comparable to $\Omega_m$ is deeply
puzzling; this is the notorious coincidence problem that has been much
discussed in the recent literature.  The only plausible explanation
that has so far been suggested is that $\Omega_\Lambda$ is a
stochastic variable and that the coincidence is due to anthropic
selection effects.  Anthropic bounds on the cosmological constant
derived in \cite{Davies,BT,LindeLambda,Weinberg87} were followed by
anthropic predictions \cite{Vilenkin95a,Efstathiou95,Martel98,GV}
suggesting values 
not far from the presently observed dark energy density.
Although controversial,
such anthropic arguments have been bolstered by the discovery of
mechanisms that may be capable of creating ensembles with different parameter
values in the context of both cosmic inflation
\cite{AV83,Linde86,Linde90} and string theory
\cite{Bousso,Banks,Donoghue01,Susskind03}, and have been applied to
other physical parameters as well
\cite{CarrRees,Rees79,Linde88a,Linde88b,Linde95,Bellido95,Vilenkin95d,Vilenkin97,Q,dimensions,t98,Agrawal,Donoghue98,Tanaka,Mario,Hogan00,Aguirre,multiverse}.

Perhaps
equally puzzling are the ``coincidences'' $\Omega_D\sim \Omega_B$ and
$\Omega_B\sim \Omega_\nu$.  These three matter components are relics of
apparently unrelated processes in the early Universe, and it is very
surprising that their mass densities are comparable to one another.
The mass density of neutrinos is $\Omega_\nu=(m_\nu/94{\rm eV})h^{-2}$, 
where $m_\nu$ is the combined mass of all three neutrino flavors. 
In this paper, we will investigate the possibility that 
$m_\nu$ is a
stochastic variable taking different values in different parts of the
Universe and that the observed value is anthropically selected. 

Before delving into details, let us briefly outline the argument. 
A small increase of $m_\nu$ can have a 
large effect on
galaxy formation.  Neutrinos stream out of the potential wells created
by cold dark matter and baryons, slowing the growth of density
fluctuations.  As a result, there will be fewer galaxies (and
therefore fewer observers) in regions with larger values of $m_\nu$.
If the suppression of galaxy formation becomes important for 
$m_\nu \gtrsim {\tilde m}_\nu$, say, then
values $m_\nu \gg {\tilde m}_\nu$ will be rarely observed because the 
density of galaxies
in the corresponding regions is very low.  
Moreover, unless the underlying particle-physics model strongly skews
the neutrino mass distribution towards values near zero, 
values $m_\nu \ll {\tilde m}_\nu$ are also unlikely to be observed,
simply because the corresponding range of $m_\nu$-values is very small.
A typical observer thus expects to find $m_\nu\sim {\tilde m}_\nu$,
i.e., a mild but non-negligible suppression of galaxy formation by
neutrinos.

\section{Probability distribution for $m_\nu$}

To make the analysis quantitative, we define the probability
distribution ${\cal P}(m_\nu)dm_\nu$ as being proportional to the
number of observers in the Universe who will measure $m_\nu$ in the
interval $dm_\nu$.  This distribution can be represented as a product
\cite{Vilenkin95a} 
\beq{P} 
{\cal P}(m_\nu)={\cal P}_*(m_\nu)n_{obs}(m_\nu).
\eeq 
Here, ${\cal P}_*(m_\nu)dm_\nu$ is the prior distribution, which
is proportional to the comoving volume of those parts of the Universe
where $m_\nu$ takes values in the interval $dm_\nu$, and $n_{obs}(m_\nu)$ is
the number of observers that evolve per unit comoving volume with a given
value of $m_\nu$. The distribution (\ref{P}) gives the probability that a
randomly selected observer is located in a region where the sum of the
three neutrino masses is in the interval $dm_\nu$.

Of course we have no idea how to estimate $n_{obs}$, but what comes to
the rescue is the fact that the value of $m_\nu$ does not directly
affect the physics and chemistry of life. As a rough approximation, we
therefore assume that $n_{obs}(m_\nu)$ is simply proportional to the
fraction of all baryons that form stars, which we approximate by
the fraction $F_M(m_\nu)$ of all matter that collapses into 
galaxy-scale haloes
(with mass greater than $M=10^{12} M_\odot$),
\beq{FM}
n_{obs}(m_\nu)\propto F_M(m_\nu).
\eeq
The idea is that there is some average number of stars per unit mass in a
galaxy and some average number of observers per star. 
The choice of the halo mass scale is based on the empirical fact that most stars
are observed to be in large halos.

The prior distribution ${\cal P}_*(m_\nu)$ depends on the extension of
the particle physics model which allows neutrino masses to vary and perhaps on
stochastic processes during inflation which randomize these
variable masses.  Some candidate prior distributions will be
discussed in Section III.

The fraction of collapsed matter $F_M(m_\nu)$ can be approximated
using the standard
Press-Schechter formalism \cite{PressSchechter}.  We assume a Gaussian density
fluctuation field $\delta({\bf x},t)$ with a variance $\sigma(M,t)$ on
the galactic scale $(M=10^{12}\Ms)$, 
\beqq
P(\delta,t)\propto \exp\left[-{\delta^2\over{2\sigma(t)^2}}\right].  
\eeq 
A collapsed halo is assumed to form when the linearized density
contrast $\delta$ exceeds  
a critical value $\delta_c$ determined by the spherical collapse model.
As detailed in Appendix A, this corresponds to 
$\delta_c\approx 1.69$ around the present epoch and $\delta_c\approx 1.63$ in the infinite future \cite{Weinberg87}.
Using the Press-Schechter approximation, we obtain
\beq{nG}
F_M(t)\propto P[\delta>\delta_c]
=\int_{\delta_c}^\infty 
P(\delta,t) d\delta = \erfc\left[{\delta_c\over\sqrt{2}\sigma_M(t)}\right].
\eeq
The collapsed fraction $F_M$ thus grows over time as the rms density
fluctuations $\sigma$ increase.

Let us now quantify the effect of neutrino masses on this process. 
For the small scale $M$ that we are considering, assuming a flat Universe,  
this fluctuation growth is well approximated by
\beq{sigmaMeq}
\sigmaM(x)\approx \left[1+{3\over 2}\Avac(\fn)\Gl(x)\right]^{p(\fn)}\sigmaM(0),
\eeq
as shown in Appendix A. The functions $\Avac$, $\Gl$ and $p$ are defined below. 
Here we have replaced $t$ by a new time variable
\beq{xEq}
x\equiv{\rhol\over\rhom} = {\Ol\over(1+z)^3\Om} = {1-\Om\over\Om}(1+z)^{-3},
\eeq
\ie, the dark-energy-to matter density ratio --- we will consider several values of $x$ below,
corresponding to the infinite future ($x=\infty$), the present epoch ($x=7/3$, our default value) and redshift unity
($x=7/24$).
The function 
\beq{GlambdaFitEq}
\Gl(x)\approx x^{1/3}\left[1+\left({x\over G_\infty^3}\right)^\alpha\right]^{-1/3\alpha},
\eeq
where $\alpha=0.795$ and
\beq{GmaxEq}
G_\infty\equiv {5\Gamma\left({2\over 3}\right)\Gamma\left({5\over 6}\right)\over 3\sqrt{\pi}}\approx 1.43728,
\eeq
describes how in the absence of massive neutrinos, 
fluctuations grow as the cosmic scale factor $a$ as long as dark energy is negligible 
($\Gl(x)\approx x^{1/3}\propto a\propto (1+z)^{-1}$ for $x\ll 1$) and then asymptote to 
a constant value as $t\to\infty$ and dark energy dominates ($\Gl(x)\to G_\infty$ as $x\to\infty$).

We are considering the case where $\mn$ varies from place to place whereas
the physics that determined the amount of baryons and cold dark matter per photon is the
same everywhere, so 
the neutrino fraction $\fn$ is given by 
\beq{fnEq}
\fn\equiv {\Omega_\nu\over\Om} 
= {\rhon\over\rhobc+\rhon} 
= \left[1+{\rhobc\over\rhon}\right]^{-1}
= \left[1+{\mbc\over\mn/3}\right]^{-1},
\eeq
where $\rhobc$ denotes the non-neutrino density, \ie, that of baryons and cold dark matter,
and $\mbc\approx (4.75\pm 0.30)$eV 
gives the measured amount of such matter per neutrino.
In other words, increasing the neutrino mass from zero will 
increase the total matter density per photon 
by a factor $\rhom/\rhobc=(1-\fn)^{-1}$.

\beq{AvacEq}
\Avac(\fn)\equiv\xeq^{-1/3} = \left({\Om\over\Ol}\right)^{1/3}(1+\zeq)
\eeq
is the factor by which the Universe has expanded between 
matter-radiation equality at $x=\xeq$ (when fluctuations effectively start to grow) 
and dark energy domination at $x=1$ (when fluctuations gradually stop growing).
Since massive neutrinos boost the matter density by a factor $(1-\fn)^{-1}$,
they delay vacuum domination until the scale factor is
larger by a factor $(1+\fn)^{1/3}$
and also, in the approximation that neutrinos are nonrelativistic
at the  matter-radiation equality epoch (valid for $\mn\simgt 1$ eV), 
cause matter-radiation equality to occur occurs earlier,
when the scale factor is smaller by a factor $(1+\fn)$.
We thus have 
\beq{Avac0Eq}
\Avac(\fn)=(1-\fn)^{-4/3}\Avac(0).
\eeq
Finally, neutrinos with nonzero mass suppress the galaxy density through the exponent 
$p(\fn)$ in \eq{sigmaMeq}, which is given by \cite{Bond80}
\beq{peq}
p(\fn) = {\sqrt{25-24\fn}-1\over 4}\approx (1-\fn)^{3/5},
\eeq
and drops from unity for $\fn=0$ to smaller values as $\fn$ increases. 

In summary, \eq{sigmaMeq} shows that the galaxy fluctuation evolution $\sigmaM(x)$ depends on the
three cosmological parameters $\Avac(0)$, $\sigmaM(0)$ and $\fn$.
To study the galaxy density as a function of neutrino fraction $\fn$  using \eq{nG},
we thus need to measure $\Avac(0)$ and $\sigmaM(0)$ from observational data without making 
any assumptions about $\fn$. We cannot do this using the values of $\Om$ and $\zeq$ reported
by, say, the WMAP team \cite{Spergel03}, since these assume that $\fn=0$ in our part of the Universe;
if $\fn>0$ here, then matter-radiation equality occurred earlier. 
We therefore repeat the Monte Carlo Markov Chain analysis reported in column 5 of Table 3 in \cite{sdsspars}, 
measuring 
$\Avac(0)$ and $\sigmaM(0)$ from the WMAP microwave background power spectrum \cite{Hinshaw03}
combined with the
Sloan Digital Sky Survey (SDSS) galaxy power spectrum \cite{sdsspower}. These measurements are independent of $\fn$ 
since this is a free
parameter in the analysis and therefore effectively marginalized over. 
This gives $\Avac(0)=3057\pm 502$,
$\sigmaM(0)\approx 0.000579\pm 0.000064$\footnote{
As our galactic scale, we take $M=10^{12}\Ms$, 
specifically a top-hat smoothing scale of $R=1.3h^{-1}$Mpc.
This corresponds to length scales about 100 times smaller than the 
matter-radiation equality scale where the matter power spectrum turns over.
}.
The above-mentioned $\mbc$-value was measured using this same MCMC analysis. 
We will use the central values for our main analysis and quantify the effect 
of the uncertainties in the discussion section.
\Eq{sigmaMeq} thus shows that for $\fn=0$, fluctuations grow 
by a factor $1+1.5\Avac(0)\Gl(x)\approx 4700$ 
by the present epoch,
which we take to be $x=\Ol/\Om\approx 0.7/0.3\approx 2.3$,
giving $\sigmaM\approx 2.7$. 
In the infinite future $x\to\infty$, fluctuations will have grown by
a factor $6600$, 
giving $\sigmaM\approx 3.8$. 
The basic reason that neutrinos have such a dramatic effect is that
these growth factors are so large, implying that even a modest change in the exponent $p(\fn)$
makes a large difference.
Taylor expanding \eq{sigmaMeq} in $\fn$ gives
\beq{gammaEq}
\sigmaM(x,\fn)\approx\sigmaM(x,0) e^{-\gamma(x)\fn}
\eeq
for $\fn\ll 1$, where $\gamma(x)\equiv 0.6\ln[1+1.5\Avac(0)\Gl(x)]-4/3 \approx 3.7$  
for the present epoch and
$\gamma\approx 3.9$ for the infinite future. 
Although \eq{gammaEq} is quite a crude approximation, underestimating the suppression,
it shows that small changes in $\Avac$ or $x$ are unimportant since they affect this exponential 
fluctuation suppression only logarithmically.

The effect of neutrino free streaming on the galactic density is
illustrated in Fig.~1 (top), which shows that the suppression is
non-negligible already for $m_\nu\sim 1$~eV. We use \eq{sigmaMeq}
in our calculations for the plots
--- the approximation \eq{gammaEq} was merely to provide qualitative intuition for the
effect.

\begin{figure}[tb]
\centerline{\epsfxsize=9.0cm\epsffile{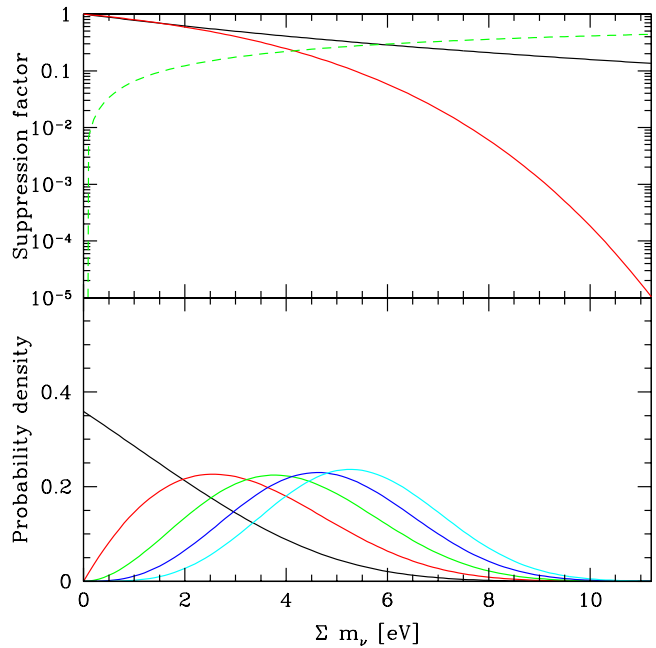}}
\caption{\label{fig1}\footnotesize 
The upper panel shows the factor by which the neutrino fraction $f_\nu$
(dashed curve) suppresses the current fluctuation
amplitude $\sigmaM$ (upper solid curve) and consequently the 
galaxy number density $n_G$ (lower solid curve).
The lower panel shows the resulting 
probability distribution for the neutrino mass sum 
for priors $m_\nu^n$ with $n=0$, 1, 2, 3 and 4, peaking from left to
right, respectively.
}
\end{figure}

%

The probability distribution ${\cal P}(m_\nu)$ is shown in Fig.~1
(bottom) for power law priors
\beqq
{\cal P}_*(m_\nu)\propto m_\nu^n,
\label{Pm}
\eeq
with $n$ ranging from 0 to 4.  For $n\geq 1$, these distributions are peaked
at $m_\nu\gtrsim 2$~eV, while in the case of a flat prior, ${\cal
P}_*(m_\nu) = {\rm const}$, the expected values are 
$m_\nu\sim 1$~eV.  This is also seen in Fig.~2, where the
distribution for a flat prior is shown using a logarithmic scale for
$m_\nu$.  

\begin{figure}[tb]
\vskip-1.3cm
\centerline{\epsfxsize=9.0cm\epsffile{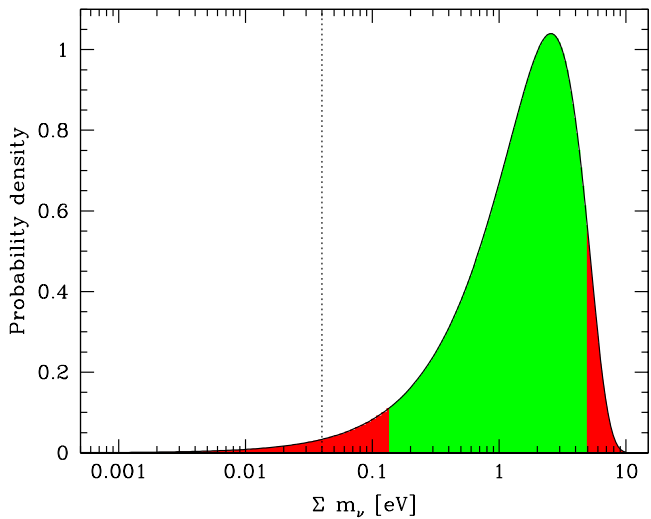}}
\vskip-0.8cm
\caption{\label{fig2}\footnotesize 
The probability distribution for the neutrino mass sum for flat prior
$(n=0)$.
The dark/red
tails contain 5\% probability each.  The dotted line shows the lower
limit 0.05 eV from atmospheric neutrino oscillations \protect\cite{Fukuda99,Kearns02,Bahcall03,King03}.
}
\end{figure}

In this discussion, we have assumed that $f_\nu\ll 1$, that is,
$m_\nu\ll 10$ eV.
Very heavy neutrinos (with $m_\nu \gg 1$ MeV) would annihilate
well before nucleosynthesis and cause no problems with structure formation.
If {\it all} neutrinos were heavy, neutrons would be stable, leading to
an equal number of protons and neutrons. As a result, most of the matter 
would end up in helium instead of hydrogen.
This lack of hydrogen would clearly suppress $n_{obs}(m_\nu)$ for 
observers like us who rely on long-lived (hydrogen burning) stars and water-based chemistry.
%
Moreover, heavy neutrinos would not be able to blow off the
envelope in supernova explosions. This means that heavy elements
formed in stellar interiors would not be dispersed to form planets
and observers.
The possibility of the electron neutrino being light
and one or two others very heavy is allowed anthropically, but it is
already ruled out by the neutrino oscillation experiments, which
constrain the mass differences to be within $0.05$ eV.

For $m_\nu\gg 100$ eV (but $\ll 1$ MeV), keeping all other physical 
parameters fixed, neutrinos would have sufficiently low thermal
velocities to act approximately as cold dark matter, thereby allowing
galaxy-size halos to form. However, the baryon fraction in these
halos would be strongly diluted, and it is therefore far from clear
whether they would be able to cool and form  
self-gravitating baryonic disks, let alone stars or observers,
with an efficiency comparable to that in our observable
universe. In other words, the calculation of anthropic constraints on
very large 
neutrino masses becomes essentially equivalent to the calculation of an
anthropic upper bound on the dark matter abundance.
We will not attempt to address this issue here, but simply
assume that the number of observers $n_{obs}(m_\nu)$  
is strongly suppressed for $m_\nu\gg 10$ eV.

We have also assumed that there are $N_\nu=3$ stable neutrinos.
Generalizing our result to $N_\nu>3$ is straightforward: as long as 
the masses are low enough for neutrino infall to be negligible, the galaxy number
density depends only on the total neutrino mass density, which for standard neutrino 
freezeout is proportional to the sum of the neutrino masses.
If the neutrinos are unstable on cosmological timescales, they suppress 
fluctuation growth only before decaying, with their decay radiation redshifting away 
to gravitationally negligible levels within a few expansion times.

\section{Prior distribution}

Following \cite{DV}, we shall now discuss possible modifications of
the standard particle physics model that could make neutrino masses
variable. 
For early work on how masses of elementary particles can vary randomly
in the context of stochastic gauge theories, see \cite{Nielsen1,Nielsen2,Nielsen3}.

Dirac-type neutrino masses can be generated
if the Standard Model neutrinos $\nu^\alpha$ mix through the Higgs
doublet VEV $\Phi$ to some gauge-singlet fermions $\nu_c^\beta$,
\beqq
g_{\alpha\beta}\Phi{\bar\nu}^\alpha\nu_c^\beta.
\eeq
The couplings $g_{\alpha\beta}$ will generally be variable in string
theory inspired models involving antisymmetric form fields $F_a$
interacting with branes. (Here, the index $a$ labels different form
fields.) $F_a$ changes its value by $\Delta F_a=q_a$ across a brane,
where $q_a$ is the brane charge. In the low-energy effective theory,
the Yukawa couplings $g_{\alpha\beta}$ become functions of the form
fields,
\beqq
g_{\alpha\beta}=g_{\alpha\beta}^{(0)}+\sum g_{\alpha\beta}^a
{F_a\over{M_p^2}} + ... ,
\label{g}
\eeq
where the summation is over all form fields, the coefficients
$g_{\alpha\beta}^{(0)}$, $g_{\alpha\beta}^a$ are assumed to be 
numbers $\sim 1$, and $M_p$ is the effective cutoff scale, which we 
assume to be the Planck mass.

In such models, closed brane bubbles nucleate and expand during
inflation \cite{Brown}, creating exponentially large regions with 
different values
of the neutrino masses. When $F_a$ changes in increments of $q_a$,
$m_\nu$ changes in increments of 
\beqq
\Delta m_\nu\sim \Phi q_a/M_p^2.
\label{Deltam}
\eeq
To be able to account for neutrino masses $\lesssim 1$ eV, we have to
require that $\Delta m_\nu\lesssim 1$ eV, that is, 
\beqq
q_a\lesssim 10^{-11} M_p^2, 
\label{smallq}
\eeq
for at least some of the brane charges. Such small values of the
charges can be achieved using the mechanisms discussed in
\cite{Bousso,DV01,Banks,Feng}. 

It should be noted that the Higgs potential and the Higgs expectation
value $\Phi$ will also generally depend on $F_a$. Moreover, each field
$F_a$ contributes a term $F_a^2/2$ to the vacuum energy density
$\rho_\L$, and regions with different values of $F_a$ will generally
have different values of $\rho_\L$. However, in the presence of
several form fields with sufficiently small charges, variations of all
these parameters are not necessarily correlated, and here we shall
assume that there is enough form fields to allow independent variation
of the relevant parameters. We can then consider a sub-ensemble of
regions with $m_\nu$ variable and all other parameters fixed. The
probability distribution for $m_\nu$ that we calculated in Section II
corresponds to such a sub-ensemble.

Let us now turn to the prior distribution ${\cal P}_*(m_\nu)$.  The
natural range of variation of $F_a$ in Eq.~(\ref{g}) is the Planck
scale, and the corresponding range of the neutrino masses is $0\leq
m_\nu^{(i)}\lesssim\Phi$. (Here, the index $i$ labels the three
neutrino mass matrix eigenvalues.) Only a small fraction of this range
corresponds to values of anthropic interest, $m_\nu\lesssim 10$~eV.
In this narrow anthropic range, we expect that the probability
distribution for $F_a$ after inflation is nearly flat 
\cite{GV01},\footnote{
A very different model for the prior distribution
was considered by Rubakov and Shaposhnikov \cite{RS89}. They assumed
that ${\cal P}_{prior}(X)$ is a sharply peaked function with a peak
outside the anthropic range ${\cal A}$ and argued that the observed
value of $X$ should then be very close to the boundary of ${\cal
A}$. We note that this is unlikely to be the case for the neutrino mass, 
since it is observed to be comfortably inside the anthropically allowed range.
If the model of \cite{RS89} applied, the peak of the full distribution would
most likely be in a life-hostile environment, where both ${\cal
P}_{prior}(X)$ and $n_{obs}(X)$ are very small. In the case of the
neutrino mass, this would mean that the number density of
galaxies is very low. This is not the case in our observable region,
indicating that the model of \cite{RS89} does not apply.}
\beqq
d{\cal P}_*= {\rm const}\cdot dF_1 dF_2 ... ,
\eeq
and that the the functions $g_{\alpha\beta}(F_a)$ are well
approximated by linear functions. If all three neutrino masses vary
independently, this implies that
\beqq
d{\cal P}_* ={\rm const}\cdot
dm_\nu^{(e)}dm_\nu^{(\mu)}dm_\nu^{(\tau)}.
\eeq
The probability for the combined mass $m_\nu=\sum m_\nu^{(i)}$ 
to be between $m_\nu$ and $m_\nu+dm_\nu$ is then proportional to the 
volume of the triangular slab of thickness $\sim dm_\nu$ in the
3-dimensional mass space,
\beqq
d{\cal P}_*\propto m_\nu^2 dm_\nu.
\label{n=2}
\eeq
Alternatively, the neutrino masses can be related to one another, for
example, by a spontaneously broken family symmetry.  If all three
masses are proportional to a single variable mass parameter, then we
expect
\beqq
d{\cal P}_*\propto  dm_\nu.
\label{n=0}
\eeq

Let us now assess how well the predictions derived from the prior
distributions (\ref{n=2}) and (\ref{n=0}) agree with observations.  We
first consider the distribution (\ref{n=2}), corresponding to
independently varying neutrino masses.  The most probable value of
$m_\nu$ for $n=2$ is $m_\nu\sim 3$~eV, and we expect both the neutrino
masses and mass differences to be $\sim 1$~eV.  This expectation,
however, is in conflict both with neutrino oscillation experiments
suggesting $\Delta m_\nu\lesssim 0.05$~eV \cite{Fukuda99,Kearns02,Bahcall03,King03} and with
astrophysical bounds \cite{Spergel03,sdsspars} which indicate a combined mass
$m_\nu\lesssim 1$~eV.

For a flat prior distribution (\ref{n=0}), the most probable value is
$m_\nu\sim 1$~eV.  If $m_\nu$ is close to this value, then the three
neutrino masses must be nearly degenerate, with $\Delta m_\nu\ll
m_\nu$.  This could be interpreted as a sign of a family symmetry.  A
90\% confidence level prediction for $m_\nu$ based on this
distribution can be obtained as outlined in Section II. This gives
\beqq
0.1~{\rm eV}<m_\nu<5~{\rm eV}.
\label{rangem}
\eeq 
The lower bound in (\ref{rangem}) is somewhat stronger than the bound
from the neutrino oscillation data \cite{Fukuda99,Kearns02,Bahcall03,King03} ($m_\nu\simgt
0.05~{\rm eV}$), while the upper bound is
somewhat weaker than the current
astrophysical bounds (\eg, 
\cite{Spergel03,Hannestad0303076,ElgaroyLahav0303089,BashinskySeljak03,Hannestad0310133,sdsspars}). 
Note that the strength of current astrophysical bounds is limited 
not by statistical errors but by systematic uncertainties in non-CMB data.
For instance, the recently claimed evidence for $\mn>0$ \cite{Allen0306386}
may result from underestimated galaxy cluster modeling uncertainties.

We finally mention the possibility that the right-handed neutrinos
$\nu_c^\alpha$ have a large Majorana mass $M_R\gg \Phi$. In this case,
small neutrino masses can be generated through the see-saw mechanism,
\beqq
m_\nu\sim g^2\Phi^2/M_R. 
\label{Majorana}
\eeq
If $M_R$ is variable, say, 
within a range $M_R\lesssim M_p$, then its most probable values are
likely to be $\sim M_p$, and the prior distribution will be peaked at
very small values of $m_\nu\sim 10^{-6}$~eV. 

This discussion suggests that the most promising scenario with
variable neutrino masses is the one with Dirac-type masses determined
by a single variable mass parameter. It yields a flat prior
distribution for $m_\nu$, Eq.~(\ref{n=0}), and the prediction
(\ref{rangem}) at 90\% confidence level.


After we submitted the original version of this paper, 
Jaume Garriga pointed out to us that see-saw-type models can
yield cosmologically interesting prior distributions for $m_\nu$ if
the Majorana mass is restricted to the range $M_R<M_R^{(max)}\ll
M_p$. Assuming first that $M_R$ is a fixed constant, while $g$ is
variable, Eq.~(\ref{Majorana}) yields the distribution
\beqq
d{\cal P}_{prior}\propto dg\propto m^{-1/2}dm.
\label{maj}
\eeq
This would give a somewhat smaller predicted neutrino mass than the
distribution (\ref{n=0}) that we used in most of our calculations. 

The distribution (\ref{maj}) applies up to $m_{max}\sim
\Phi^2/M_R$. In order to have $m_{max}\gtrsim 0.1$ eV, we need
$M_R\lesssim 10^{13}$ GeV.

If both $M_R$ and $g$ are variable, then, assuming a flat prior for
$M_R$, Eq.~(\ref{maj}) still applies, but now $m_{max}\sim
\Phi^2/M_R^{(max)}$, so we need $M_R^{(max)}\lesssim 10^{13}$ GeV.
An attractive feature of this scenario is that the increment of
$m_\nu$ in Eq.(\ref{Deltam}) gets suppressed by an additional factor
$\Phi/M_R$, and Eq.(\ref{smallq}) gets replaced by a much weaker
constraint $q_a/M_p^2 \lesssim (M_R/10^{13}~{\rm GeV})^{-1}$.

\section{Discussion}

\begin{figure}[tb]
\vskip-1.3cm
\centerline{\epsfxsize=9.0cm\epsffile{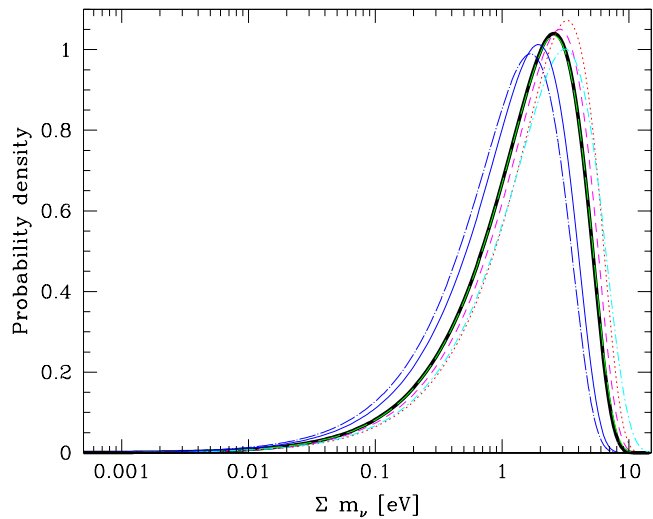}}
\vskip-0.8cm
\caption{\label{fig3}\footnotesize 
Same as Figure 2, but showing the robustness of the
results to changing various assumptions.
%
We have changed the baseline calculation from Figure 2
(heavy black curve) by evaluating the galaxy density in the
infinite future $x\to\infty$ (dotted red/grey curve)
and at redshift unity, $x=7/3(1+z)^3=7/24$ (dot-long-dashed blue/grey curve), 
decreasing the density threshold to $\delta_c=1.5$ (short-dashed magenta/grey curve),
lowering the primordial fluctuation amplitude on the galactic scale by 25\% (solid blue/grey curve),
including the baryon correction as per Eisenstein \& Hu (long-dashed green curve)
and 
additionally including the neutrino infall correction as per Eisenstein \& Hu 
(dot-dashed cyan/grey curve).
%
}
\end{figure}

In conclusion, we have found that the small values of the neutrino masses may be due to
anthropic selection. If so, then the most promising model appears to
be the one with a flat prior distribution, ${\cal P}_*(m_\nu)={\rm
const}$.  The range of $m_\nu$ predicted in this model,
Eq.~(\ref{rangem}), has interesting implications for both particle
physics and cosmology. On the particle physics side, neutrino masses
in this range are nearly degenerate, suggesting extensions of the
Standard Model involving a spontaneously broken family symmetry. On
the cosmological side, a combined neutrino mass of $\gtrsim 1$~eV has
a non-negligible effect on galaxy formation. This means that it must
be taken into account in precision tests of inflation that measure the
shape of the primordial power spectrum by combining microwave
background and large-scale structure data.

Let us close by discussing the importance of the assumptions we have made and 
outlining some open problems for future work. The
purpose of this brief paper and the prediction of \eq{rangem} is
merely to demonstrate that anthropic selection effects {\it may} be
able to explain the neutrino masses, and much work needs to be done to
place this hypothesis on a firmer footing.

\subsection{Robustness to approximations and measurement errors}

To quantify the robustness of our results, Figure 3 shows how the probability distribution
for $m_\nu$ changes 
when various assumptions are altered. 

First of all, the parameters $\Avac$ and $\sigmaM(0)$ 
that we used have non-negligible measurement uncertainties. 
We see that lowering $\sigmaM(0)$ by 25\% (by about twice its measurement
uncertainty) lowers the $\mn$-prediction slightly. Changing $\Avac$ by within its observational
uncertainty has an even weaker effect since, 
as we saw, it enters only logarithmically. Altering the galactic scale $M$ affects $\sigma_M$ and hence the results
only weakly, because of the flatness of the dimensionless power spectrum $k^3 P(k)$ on galactic scales.

Second, our calculations involved various approximations.  We used the
Press-Schechter approximation with density threshold $\delta_c\approx 1.69$
as per Appendix A; lowering this to account for post-virialization infall 
as discussed in \cite{Martel98} is seen to raise the
$\mn$-prediction slightly. Our fluctuation growth treatment of
\eq{sigmaMeq} is highly accurate in the limit of small mass scales $M$
and low baryon fraction $\Omega_b/\Omega_m\ll 1$, agreeing with a
CMBfast \cite{cmbfast} numerical calculation to within a few
percent. Figure 3 shows that for the observed baryon fraction
$\Omega_b/\Omega_m\approx 0.15$, switching to an exact treatment of
baryon effects makes virtually no difference.
The
cosmic expansion eventually slows neutrinos enough for them to start
clustering on galaxy scales, and if this happens before dark energy
domination, then it reduces $\gamma$ \eq{gammaEq}. Since a small
fraction of the neutrinos will be in the low tail of their velocity
distribution, there is a slight infall correction even for the low
$\mn$-range we have considered, and Figure 3 shows that this increases
our $\mn$-prediction slightly. Finally, we have used the cutoff value
of $x\approx 7/3$, which amounts to using the reference class of
observers in galaxies that have formed by now.  Figure 3 shows that if
we ask instead what would be observed from a random galaxy among all
galaxies that ever form (setting $x=\infty$), then the
$\mn$-prediction increases slightly.  Conversely, it also shows that considering
only observers in galaxies that formed by redshift unity 
decreases 
the $\mn$-prediction. In conclusion, Figure 3 shows that
although many of these assumptions make marginal differences, none of
them affect the qualitative conclusions, since they all shift the
predicted probability distribution by much less than one standard
deviation.


\subsection{Effects of other parameters}

The standard models of cosmology and particle physics involve of order
10 and 28 free parameters, respectively. In order to apply anthropic
constraints to them, it is crucial to know both which of them can
vary, and what the interdependencies or correlations between them
are. It is likely that at least some of the cosmological parameters
(the baryon-to-photon ratio, say, via baryogenesis) are determined by
particle physics parameters in a way that we have yet to understand,
and many particle physics parameters may in turn be determined by a
smaller number of parameters or vacuum expectation values of some
deeper underlying theory. A proper analysis of anthropic predictions
should therefore be done in the multi-dimensional space spanned by all
fundamental variable parameters.


Such correlations between parameters must ultimately be taken into
account not only for computing the theoretical prior ${\cal P}_*$ of
\eq{P}, but also when computing the factor $n_{obs}$ in this equation,
which incorporates the observational selection effect. The reason is
that strong degeneracies are present which can in many cases offset a
detrimental change in one parameter by changes in others. For
instance, suppressed galaxy formation caused by increased $m_\nu$ can to some extent be compensated by
decreasing $\rho_\Lambda$, by increasing the dark-matter-to-photon ratio or by increasing the
CMB fluctuation amplitude $Q$ above the value $\sim 10^{-5}$ that we
observe \cite{Q,Aguirre} --- if any of these three parameters can vary, that is. In the present paper,
we have merely considered the simple case where all relevant
parameters except $m_\nu$ 
(\ie, the comoving densities of baryons and dark matter, 
the physical density of dark energy, and
the fluctuation amplitude $Q$)
are kept fixed at their observed values,
with no account for scatter due to variation across an ensemble or
from measurement uncertainties.
A more detailed study of this issue is given in \cite{anthrolambdanu} 
and shows that our present results for $\fn$ are rather robust to assumptions 
about $\rhol$.


This is closely related to the issue of how much information one
wishes to include in the factor $n_{obs}$ in \eq{P} \cite{Bostrom,Aguirre}. One
extreme is including only the existence of observers, the other
extreme is including all available knowledge (even, say, experimental
constraints on $m_\nu$). 
As one includes more such information, the anthropic factor becomes
progressively less important, and the calculation acquires the flavor
of a prediction rather than an explanation.  In the context of a
multiparameter analysis, the question is whether to use the measured
values of other parameters (in our case non-neutrino parameters) or
marginalize over them.  Our fixing non-neutrino parameters at their
observed values is therefore equivalent to factoring in the
information from the measurements of these parameters.%


\bigskip

Arguably the most interesting outstanding question is 
whether the fundamental equations that govern our Universe
do or do not allow physical quantities such as the neutrino masses to vary from place to place.
Calculations of anthropic selection effects may prove useful for shedding light on this.
In any case, for quantities that do vary, the inclusion of anthropic selection effects such as the one we have 
evaluated in this paper is clearly not optional when calculating what the
theory predicts that we should observe.

\bigskip
We thank Anthony Aguirre, Gia Dvali, Jaume Garriga, Martin Rees and
Douglas Scott for helpful comments.  MT was supported by NSF grants
AST-0071213 \& AST-0134999, NASA grant NAG5-11099, by a David and
Lucile Packard Foundation fellowship and a Cottrell Scholarship from
Research Corporation. AV was supported in part by the National Science
Foundation and the John Templeton Foundation.

\appendix

\section{Growth of linear perturbations}

In this Appendix, we derive and test the approximation 
given by \eq{sigmaMeq} for how small-scale matter fluctuations grow 
in the presence of radiation, cold dark matter and neutrinos.
There are two reasons why this simple approximation complements an exact ``black box'' calculation 
with CMBfast \cite{cmbfast} or a nearly exact approximation with the Eisenstein \& Hu fitting 
software \cite{EisensteinHu99}. First, for a qualitative argument like the one we make in this paper,
it is desirable to have a simple intuitive understanding of the underlying physics that includes
only those complications that really matter for the argument.
Second, neither CMBfast nor the Eisenstein \& Hu package were designed to be valid for extreme cosmological
parameters such as those corresponding to the infinite future, and indeed break down in this limit.

\begin{figure}[tb]
\centerline{\epsfxsize=9.0cm\epsffile{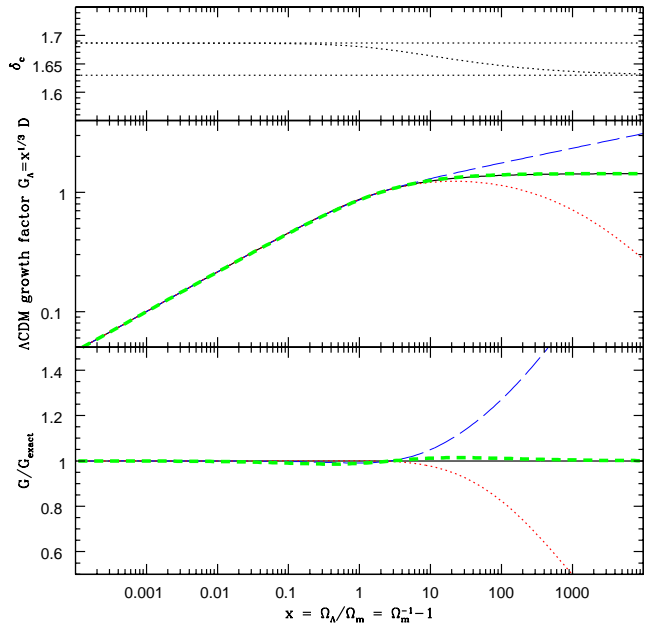}}
\caption{\label{GrowthFig}\footnotesize 
The LCDM growth function $\Gl$ is shown as a function of cosmic time, 
first growing as the scale factor $a\propto x^{1/3}$, 
then asymptoting to $1.44$ as dark energy halts the fluctuation growth at $x\simgt 1$.
The middle panel shows the exact result of \eq{GlambdaEq} (solid curve),
our approximation given by \eq{GlambdaFitEq} (green thick dashed curve),
the Carroll, Press \& Turner approximation \protect\cite{Carroll92}
(red dotted curve) and the power law approximation $\Gl=\Om^{0.21}$ \protect\cite{openreion}
(blue long-dashed curve). In the bottom panel, the various approximations have been divided by the exact result,
showing that \eq{sigmaMeq} is accurate to better than 1.5\% for all $x$.
The top panel shows the collapse density threshold $\delta_c(x)$ dropping from 
$1.6865$ early on to $1.62978$ in the infinite future.
}
\end{figure}

\subsection{The $\Lambda$CDM case}

For a flat Universe with only 
pressureless matter (dark and baryonic) and a cosmological constant,
the growth of density fluctuations is given by 
$\delta\propto\Gl(x)$, where \cite{Heath77,Martel98}
\beq{GlambdaEq}
\Gl(x)={5\over 6}\sqrt{1+{1\over x}}\int_0^x{dy\over y^{1/6}(1+y)^{3/2}}.
\eeq
We find that our $\Gl(x)$ fit 
defined by \eq{GlambdaFitEq}
is accurate to better than 1.5\% for all $x$ and becomes exact both
in the limits $x\to 0$ (when $\Gl\to x^{1/3}$) and 
$x\to\infty$ (when $\Gl\to G_\infty$).
\Fig{GrowthFig} shows that this approximation greatly improves on the standard
Carroll, Press \& Turner \cite{Carroll92} and power law fits for our present purposes, 
since these were designed to be accurate only in the past and present 
and have the wrong limiting behavior in the future 
as $x\to\infty$, $\Om\to 0$ and $\Ol\to 1$.
For flat models, 
we have $\Om=1/(1+x)$, $\Ol=x/(1+x)$ and $x=(\Om^{-1}-1)^{-1}$, so
in terms of 
the standard linear growth factor $D\equiv\Gl/x^{1/3}$, 
the three approximations shown in \Fig{GrowthFig} are
\beq{MyDapproxEq}
D\approx \left[1+\left({1-\Om\over \Om G_\infty^3}\right)^\gamma\right]^{-1/3\gamma},
\eeq
\beqa{CarrollEq}
D
&\approx&{5\over 2}\Om\left[\Om^{4/7} - \Ol + 
\left(1+{\Om\over 2}\right)\left(1+{\Ol\over 70}\right)\right]^{-1}\nonumber\\
&=&{350\Om\over 140\Om^{4/7} + (209-\Om)\Om + 2}
\eeqa
and 
\beq{PowerLawDeq}
D\approx\Om^{-0.21},
\eeq
respectively.

\subsection{Including radiation}

Early on, dark energy was negligible but radiation was gravitationally important,
causing density fluctuations to grow as
$\delta\propto\Gg(x)$, where \cite{PeeblesOpus}
\beq{GgammaEq}
\Gg(x) = 1 + {3\over 2}\left({x\over\xeq}\right)^{1/3}.
\eeq
Since both $\Gg(x)$ and $\Gl(x)$ accurately describe the
growth during the matter-dominated epoch $\xeq\ll x\ll 1$, with $\Gg\propto\Gl\propto x^{1/3}$
during this period,
we can combine them to obtain the approximation
\beq{Gtoteq}
\G(x)\approx 1+{3\over 2}\Avac\Gl(x),
\eeq
which is accurate for all $x$. Here the constant $\Avac$ is defined by \eq{AvacEq}.
In essence, fluctuations grow as $\delta\propto a\propto x^{1/3}$ between 
matter domination ($x=\xeq$) and dark energy domination ($x=1$), giving a 
net growth of $\Avac=\xeq^{-1/3}$.
\Eq{Gtoteq} shows that they grow by an extra factor of 1.5 by starting slightly before 
matter domination and by an extra factor $\Gl(\infty)\approx 1.44$ by
continuing to grow slightly after dark energy domination.

\subsection{Including neutrinos}

As shown by \cite{Bond80}, 
the result $\delta\propto a$ is generalized to $\delta\propto a^p$ 
when a fraction of the matter is clustering inert and remains spatially uniform.
The new exponent $p<1$ is given by \eq{peq} where, in 
the case that we are focusing on here, the inert fraction is the neutrino 
fraction $\fn$.\footnote{The result is more general \cite{Ma96,EisensteinHu99}, 
and applies also when the inert density components correspond to dark energy or spatial curvature.
If we let $\Om$ denote the density fraction that is not inert (that clusters), then
the approximation to \eq{peq} given by $p\approx \Om^{3/5}$ is quite accurate,
being exact both for $\Om=0$ and to first order in $(1-\Om)$ for all $1-\Om\ll 1$.
This is the familiar result that $d\ln\delta/d\ln a\approx\Om^{0.6}$. 
}
This motivates our approximation in \eq{sigmaMeq}, which simply generalizes 
\eq{Gtoteq} by introducing the neutrino-dependent exponent $p(\fn)$.

We have tested this approximation by comparing \eq{sigmaMeq} with exact results 
using the CMBfast software \cite{cmbfast} and the semianalytic approximation
of Eisenstein \& Hu \cite{EisensteinHu99}, finding excellent agreement (to within a few percent)
with both in the small-scale limit for $x\sim 1$ and negligible baryon fraction.
In the distant future limit $x\to\infty$, both CMBfast and the Eisenstein \& Hu software
break down, since they were not designed to be accurate for such unusual parameter values
($\Ol=1$, \etc). 
For the parameter ranges of interest to us, there are small corrections for the effects of 
both baryons and neutrino infall, which we quantified in \fig{fig3} in the discussion section.

\subsection{The collapse density threshold $\delta_c$}

In the top panel of \fig{GrowthFig}, we have numerically computed the 
collapse density threshold $\delta_c$ as a function of cosmic time $x$,
defined as the linear perturbation theory overdensity that a top
hat fluctuation would have had at the time when it collapses.
We see that it varies only very weakly with time (note the expanded vertical scale in the figure),
dropping from the familiar cold dark matter value 
$\delta_c(0)=(3/20)(12\pi)^{2/3}\approx 1.68647$ early on
to the limit 
$\delta_c(\infty)= (9/5) 2^{-2/3} G_\infty\approx 1.62978$ \cite{Weinberg87}
in the infinite future.
This calculation neglects the effect of neutrinos.
Since their effect is to contribute a gravitationally inert component just as dark energy,
we will ignore their effect on $\delta_c(x)$, assuming that they merely cause a slight 
horizontal stretching of the curve (which is seen to be almost constant anyway).


\clearpage
  

\begin{thebibliography}{99}


\bibitem{Davies}
P.C.W. Davies and S. Unwin, Proc. Roy. Soc. {\bf 377}, 147 (1981).

\bibitem{BT}
\rfbook\nnn Barrow J D\dualand \nnn Tipler F J;1986;The 
Anthropic Cosmological Principle;Clarendon Press;Oxford

\bibitem{LindeLambda}
A.D. Linde, in {\it 300 Years of Gravitation}, ed. by S.W. Hawking and
W. Israel, Cambridge University Press, Cambridge (1987).

\bibitem{Weinberg87}
\rf\nn Weinberg S;1987;Phys. Rev. Lett.;59;2607

\bibitem{Vilenkin95a}
\rf\nn Vilenkin A;1995;Phys. Rev. Lett.;74;846

\bibitem{Efstathiou95}
\rf\nn Efstathiou G;1995;MNRAS;274;L73

\bibitem{Martel98}
\rf\nn Martel H, \nnn Shapiro P R\multiand\nn Weinberg S;1998;ApJ;492;29


\bibitem{GV}
For a recent discussion and references see J. Garriga and
A. Vilenkin, Phys. Rev. {\bf D67}, 043503 (2003).


\bibitem{AV83}
A. Vilenkin, Phys. Rev. {\bf D27}, 2848 (1983).

\bibitem{Linde86}
\rf\nnn Linde A D;1986;Phys. Lett.;175B;395

\bibitem{Linde90}
\rfbook For a review see \nnn Linde A D;1990;Particle Physics
and Inflationary Cosmology;Harwood;{Chur, Switzerland}



\bibitem{Bousso}
R. Bousso and J. Polchinski, JHEP 0006:006 (2000).

\bibitem{Banks}
T. Banks, M. Dine and L. Motl, JHEP 0101:031 (2001).

\bibitem{Donoghue01}
J.F. Donoghue, Int. J. Mod. Phys. {\bf A16S1C}, 902 (2001).

\bibitem{Susskind03}
\rfprep\nn Susskind L;2003;hep-th/0302219



\bibitem{CarrRees}
\rf\nn Carr B\dualand \nnn Rees M J;1979;Nature;278;605

\bibitem{Rees79}
\rf\nnn Rees M J;1979;Physica Scripta;21;614

\bibitem{Linde88a}
\rf\nnn Linde A D;1988;Phys. Lett. B;201;437

\bibitem{Linde88b}
\rf\nnn Linde A D\dualand \nnn Zelnikov M I;1988;Phys. Lett. B;215;59

\bibitem{Linde95}
\rf\nnn Linde A D;1995;Phys. Lett. B;351;99

\bibitem{Bellido95}
\rf\nn Garc\'\i a-Bellido J\dualand\nnn Linde A D;1995;Phys. Rev. D;251;429



\bibitem{Vilenkin95d}
A. Vilenkin, in {Cosmological constant and the evolution of the
universe}, ed by K. Sato, T. Suginohara and N. Sugiyama (Universal
Academy Press, Tokyo, 1996); gr-qc/9512031.


\bibitem{Vilenkin97}
\rf\nn Vilenkin A\dualand\nn Winitzki S;1997;Phys. Rev. D;55;548

\bibitem{Q}
\rf\nn Tegmark M\dualand\nnn Rees M J;1998;ApJ;499;526

\bibitem{dimensions}
\rf\nn Tegmark M;1997;Class. Quant. Grav.;14;L69

\bibitem{t98}
\rf\nn Tegmark M;1998;Ann. Phys.;270;1

\bibitem{Agrawal}
B. Agrawal, S.M. Barr, J.F. Donoghue and D. Seckel,
Phys. Rev. Lett. {\bf 80}, 1822 (1998); Phys. Rev. {\bf D57}, 5480
(1998). 

\bibitem{Donoghue98}
J.F. Donoghue, Phys. Rev. {\bf D57}, 5499 (1998).

\bibitem{Tanaka}
J. Garriga, T. Tanaka and A. Vilenkin, Phys. Rev. {\bf D60}, 023501
(1999).

\bibitem{Mario}
J. Garriga, M. Livio and A. Vilenkin, Phys. Rev. {\bf D61}, 023503
(2000). 

\bibitem{Hogan00}
\rf\nnn Hogan C J;2000;Rev. Mod. Phys.;72;1149
 
\bibitem{Aguirre}
\rf\nn Aguirre A;2001;PRD;64;083508
 
\bibitem{multiverse}
\rfprep\nn Tegmark M;2003;astro-ph/0302131

\bibitem{PressSchechter}
\rf\nnn Press W H\dualand\nn Schechter P;1974;ApJ;187;425

\bibitem{Bond80}
\rf\nnn Bond J R, \nn Efstathiou G\multiand\nn Silk, J;1980;PRL;45;1980

\bibitem{Spergel03}
\rf\nnn Spergel D N {\etal};2003;ApJS;148;175

\bibitem{sdsspars}
\rf\nn Tegmark M {\etal};2004;PRD;69;103501	

\bibitem{Hinshaw03}
\rf\nn Hinshaw G {\etal};2003;ApJS;148;135

\bibitem{sdsspower}
\rf\nn Tegmark M {\etal};2004;ApJ;606;702

\bibitem{RS89}
V.A. Rubakov and M.E. Shaposhnikov, Mod. Phys. Lett. {\bf A4}, 107
(1989). 

\bibitem{Fukuda99}
\rf\nn Fukuda Y {\etal};1999;Phys. Rev. Lett.;82;1810

\bibitem{Kearns02}
\rfprep\nnn Kearns E T;2002;hep-ex/0210019
 
\bibitem{Bahcall03}
\rf\nnn Bahcall J N;2003;JHEP;0311;004

\bibitem{King03}
\rf\nn King S;2004;Rept.~Prog.~Phys.;67;107



\bibitem{DV}
G. Dvali and A. Vilenkin, unpublished.

\bibitem{Nielsen1}
\rf\nnn Nielsen H B\dualand\nnn Froggatt C D;197?;Nucl. Phys B;147;277

\bibitem{Nielsen2}
\rf\nnn Nielsen H B\dualand\nnn Froggatt C D;1980;Nucl. Phys B;164;114

\bibitem{Nielsen3}
\rfbook\nnn Froggatt C D\dualand\nnn Nielsen H B;1991;Origin of Symmetries;World Scientific;Singapore

\bibitem{Brown}
J.D. Brown and C. Teitelboim, Nucl. Phys. {\bf 279}, 787 (1988).

\bibitem{DV01}
G. Dvali and A. Vilenkin, Phys. Rev. {\bf D64}, 063509 (2001). 

\bibitem{Feng}
J.L. Feng, J. March-Russell, S. Sethi and F. Wilczek, Nucl. Phys. {\bf
B602}, 307 (2001).

\bibitem{GV01}
J. Garriga and A. Vilenkin, Phys. Rev. {\bf D64}, 023517 (2001).

\bibitem{Hannestad0303076}
\rf\nn Hannestad S;2003;JCAP;05;004

\bibitem{ElgaroyLahav0303089}
\rf\nn Elgaroy O\dualand\nn Lahav O;2003;JCAP;0304;004

\bibitem{BashinskySeljak03}
\rf\nn Bashinsky S\dualand Seljak U;2004;PRD;69;083002

\bibitem{Hannestad0310133}
\rfprep\nn Hannestad S;2003;astro-ph/0310133

\bibitem{Allen0306386}
\rf\nnn Allen S W, \nnn Schmidt R W\multiand\nnn Bridle S L;2003;MNRAS;346;593
 

 

\bibitem{cmbfast}
\rf\nn Seljak U\multiand\nn Zaldarriaga M;1996;ApJ;469;437

\bibitem{anthrolambdanu}
\rf\nn Pogosian L, 
\nn Vilenkin A\multiand\nn Tegmark M;2004;JCAP;407;5

\bibitem{Bostrom}
\rfbook\nn {Bostr\"o m} N;2002;Anthropic Bias: Observation Selection Effects in Science and 
Philosophy;Routledge;{New York}




\bibitem{EisensteinHu99}
\rf\nnn Eisenstein D J\dualand\nn Hu W;1999;ApJ;511;5

\bibitem{Carroll92}
\rf\nnn Carroll S M, \nnn Press W H\multiand\nnn Turner E L;1992;ARA\&A;30;499

\bibitem{openreion}
\rf\nn Tegmark M\dualand\nn Silk J;1995;ApJ;441;458 

\bibitem{Heath77}
\rf\nnn Heath D J;1977;MNRAS;179;351

\bibitem{PeeblesOpus}
P.~J.~E.~Peebles,{\it Principles of Physical Cosmology}, Princeton University Press, 
Princeton, New Jersey (1993).

\bibitem{Ma96}
\rf\nnn Ma C P;1996;ApJ;471;13 









  

\end{thebibliography}
\end{document}